\documentclass[journal=nalefd,manuscript=article]{achemso}
\usepackage{chemformula} % Formula subscripts using \ch{}
\usepackage[T1]{fontenc} % Use modern font encodings
\usepackage{amsfonts}
\usepackage{bm}
\usepackage{multirow}
\usepackage{color}
\usepackage{multirow}
\usepackage{epsfig}
\usepackage{graphicx}
\usepackage[normalem]{ulem}
\usepackage{amsmath}
\usepackage{amssymb}
\usepackage{bm}
\usepackage{dcolumn}
\usepackage{braket}
\usepackage{bbold}
\usepackage{natbib}
\usepackage{array}
\usepackage{tabulary}
\usepackage{multirow}
\usepackage{gensymb}
\usepackage{ragged2e}
\usepackage{setspace}
\usepackage{indentfirst}

\usepackage{array}
\usepackage{tabulary}
\usepackage{multirow}
\newcolumntype{C}[1]{>{\centering\arraybackslash}p{#1}}
\newcolumntype{L}[1]{>{\flushleft\arraybackslash}p{#1}}
\author{Chang-Yin Ji}
\affiliation
{Centre for Quantum Physics, Key Laboratory of Advanced Optoelectronic Quantum Architecture and Measurement (MOE), School of Physics, Beijing Institute of Technology, Beijing, 100081, China}
\alsoaffiliation
{Beijing Key Lab of Nanophotonics \& Ultrafine Optoelectronic Systems,
School of Physics, Beijing Institute of Technology, Beijing, 100081, China}
\altaffiliation{Contributed equally to this work}

\author{Wenze Lan}
\affiliation
{Beijing National Laboratory for Condensed Matter Physics, Institute of Physics, Chinese Academy of Sciences, Beijing 100190, China}
\alsoaffiliation
{School of Physical Sciences, CAS Key Laboratory of Vacuum Physics, University of Chinese Academy of Sciences, Beijing 100190, China}
\altaffiliation{Contributed equally to this work}

\author{Peng Fu}
\affiliation{Beijing National Laboratory for Condensed Matter Physics, Institute of Physics, Chinese Academy of Sciences, Beijing 100190, China}
\alsoaffiliation
{School of Physical Sciences, CAS Key Laboratory of Vacuum Physics, University of Chinese Academy of Sciences, Beijing 100190, China}

\author{Gang Wang}
\affiliation
{Centre for Quantum Physics, Key Laboratory of Advanced Optoelectronic Quantum Architecture and Measurement (MOE), School of Physics, Beijing Institute of Technology, Beijing, 100081, China}
\alsoaffiliation
{Beijing Key Lab of Nanophotonics \& Ultrafine Optoelectronic Systems,
School of Physics, Beijing Institute of Technology, Beijing, 100081, China}

\author{Changzhi Gu}
\affiliation{Beijing National Laboratory for Condensed Matter Physics, Institute of Physics, Chinese Academy of Sciences, Beijing 100190, China}
\alsoaffiliation
{School of Physical Sciences, CAS Key Laboratory of Vacuum Physics, University of Chinese Academy of Sciences, Beijing 100190, China}

\author{Jiafang Li}
\affiliation
{Centre for Quantum Physics, Key Laboratory of Advanced Optoelectronic Quantum Architecture and Measurement (MOE), School of Physics, Beijing Institute of Technology, Beijing, 100081, China}
\alsoaffiliation
{Beijing Key Lab of Nanophotonics \& Ultrafine Optoelectronic Systems,
School of Physics, Beijing Institute of Technology, Beijing, 100081, China}
\email{jiafangli@bit.edu.cn}

\author{Yugui Yao}
\affiliation
{Centre for Quantum Physics, Key Laboratory of Advanced Optoelectronic Quantum Architecture and Measurement (MOE), School of Physics, Beijing Institute of Technology, Beijing, 100081, China}
\alsoaffiliation
{Beijing Key Lab of Nanophotonics \& Ultrafine Optoelectronic Systems,
School of Physics, Beijing Institute of Technology, Beijing, 100081, China}
\email{ygyao@bit.edu.cn}

\author{Baoli Liu}
\email{blliu@iphy.ac.cn}
\affiliation{Beijing National Laboratory for Condensed Matter Physics, Institute of Physics, Chinese Academy of Sciences, Beijing 100190, China}
\alsoaffiliation
{CAS Center for Excellence in Topological Quantum Computation, CAS Key Laboratory of Vacuum Physics, University of Chinese Academy of Sciences, Beijing 100190, China}
\alsoaffiliation
{Songshan Lake Materials Laboratory, Dongguan, Guangdong 523808, China}

\title{Probing Phase Transition of Band Topology via Radiation Topology}

\begin{document}
	%%%%%%%%%%%%%%%%%%%%%%%%%%%%%%%%%%%%%%%%%%%%%%%%%%%%%%%%%%%%%%%%%%%%%
	%% The "entry" environment can be used to create an entry for the
	%\begin{tocentry}
	
	%\includegraphics[width=9 cm,height=3.5 cm]{TOC}
	%Some journals require a graphical entry for the Table of Contents.
	%This should be laid out ``print ready'' so that the sizing of the
	
	%
	%Inside the \texttt{tocentry} environment, the font used is Helvetica
	%8\,pt, as required by \emph{Journal of the American Chemical
	%The surrounding frame is 9\,cm by 3.5\,cm, which is the maximum
	%permitted for  \emph{Journal of the American Chemical Society}
	%graphical table of content entries. The box will not resize if the
	%content is too big: instead it will overflow the edge of the box.
	%
	%This box and the associated title will always be printed on a
	%separate page at the end of the document.
	%
	%\end{tocentry}
	

\begin{abstract}
Topological photonics has received extensive attention from researchers because it provides brand new physical principles to manipulate light. Band topology of optical materials is characterized using the Berry phase defined by Bloch states. Until now, the criteria for experimentally probing the topological phase transition of band topology has always been relatively lacking in topological physics. Moreover, radiation topology can be aroused by the far-field polarizations of the radiating Bloch states, which is described by the Stokes phase. Although such two types of topologies are both related to Bloch states on the band structure, it is rather surprising that their development is almost independent. Here,  we reveal that the phase transition of band topology can be probed by the radiation topology. We theoretically design and experimentally demonstrate such an intriguing phenomenon by constructing photonic crystals that support optical analogs of quantum spin Hall effects. The results show that the topological charge of the far-field polarization vortex changes from +1 to -2 or from -2 to +1 when the band topology changes from trivial to non-trivial, which provides a new criterion to probe the phase transition of band topology using radiation topology. Our findings not only provide an insightful understanding of band topology and radiation topology, but also can serve as a novel route to manipulate the near and far fields of light.

\end{abstract}
	
	\newpage

\section{Introduction}
Topological photonics is an emerging field that provides whole new research perspectives to manipulate the flow of light \cite{lu2014topological,ozawa2019topological,tang2022topological,
yin2020manipulating,liu2021topological,wang2022fundamentals}. The band topology in photonics is derived from condensed matter physics \cite{haldane2008possible}, which classifies insulators and semimetals into different types. The bulk of photonic topological insulators (PTIs) are still insulating but their boundaries do conduct. Due to topological protection, topological states propagated at boundaries of the PTIs are very robust even with imperfections. Such robust unidirectional transport characteristics of the topological states make PTIs striking materials in the applications of optical devices, such as high transmission waveguide \cite{wang2009observation,he2019silicon}, robust optical delay lines \cite{hafezi2011robust}, topological lasers \cite{bandres2018topological,zeng2020electrically,amelio2020theory} and quantum photonic circuits \cite{chen2021topologically}. The topological quasiparticles, such as topological exciton-polaritons and phonon-polaritons, can also be obtained by the strong coupling of topological photons with exciton \cite{li2021experimental,liu2020generation} and topological photons with phonons \cite{guddala2021topological}. These aforementioned novel phenomena and applications are attributed to the topological phase transitions of band topology. Thus, it is crucial to accurately characterize the topological properties of optical materials.

In theory, we can characterize the topological property of the bulk physics and its evolution in the parameter space by calculating the topological invariant \cite{hasan2010colloquium}. This method requires obtaining the Bloch states on the band structures \cite{xiao2010berry}, which makes it very difficult to experimentally measure the topological invariant, especially in nanophotonics. Currently, an experimentally feasible approach for probing the topological properties of band structures is to explore the existence of in-gap topological states in a strip-shaped sample based on the principle of bulk-boundary correspondence. \cite{chiu2016classification, ozawa2019topological}. However, such method usually requires a full bandgap which obviously hinders its application to more general situations. For example, topological states merge with trivial bulk states when there is no full bandgap \cite{chen2019corner,hu2021nonlinear,guo2021observation,wang2021quantum}. In this case, spectral measurements alone are not enough, the near-field distribution of the mode is needed to determine whether the state is topological. Thus, it is necessary and interesting to find new methods that investigates the band topology.

When the frequency of the Bloch state (photonic mode) is above the light cone \cite{yin2020manipulating,liu2021topological,wang2022fundamentals}, the Bloch state can leak into the far field. In such case, one can map the far-field polarization of the radiating Bloch state in the momentum space. The radiation topology can be formed when there are polarization singularities in the momentum space \cite{yin2020manipulating,liu2021topological,wang2022fundamentals}. A well-known example is the radiation topology caused by bound states in the continuum (BICs) \cite{zhen2014topological,hsu2016bound}. Recently, some works turn to investigate the dynamics of radiation topology, whose processes are accompanied by many interesting physical phenomena \cite{liu2019circularly,zeng2021dynamics,ye2020singular,jin2019topologically,kang2021merging,kang2022merging,yoda2020generation,yin2020observation}.
However, little attention has been paid to the band topological properties when studying the radiative topology of the photonic band structure, and vice versa. The band topology and radiation topology are almost viewed as independent themes in the realm of topological photonics. In fact, such two vibrant topological phenomena are both rooted in the Bloch states on the band structures in photonic crystals. Thus, it is very intriguing to ask whether we can use radiation topology to probe the phase transition of band topology. Their connections can not only add to our deep understanding of different topological branches, but also provide brand new schemes to manipulate the flow
of light at material domain walls and light radiation in the far-field.

Here, we reveal that one can probe the phase transition of band topology via radiation topology, leading to new criteria for studying band topology. To demonstrate such idea, we theoretically design and experimentally fabricate photonic crystal slabs (PhCSs) that have radiation topology and exhibit the topological phase transition of band topology in the visible range. Theoretical results show that when the band topology changes from trivial to non-trivial, the topological charge of the radiation topology changes from +1 to -2 or from -2 to +1, which is further verified by experimental measurements. Moreover, we show that the topological phase transition of band topology can also serve as a novel route toward the manipulation of polarization and quality ($Q$) factor of the radiation field on the photonic bands.

\begin{figure}
\centering{}\includegraphics[width=0.7\textwidth]{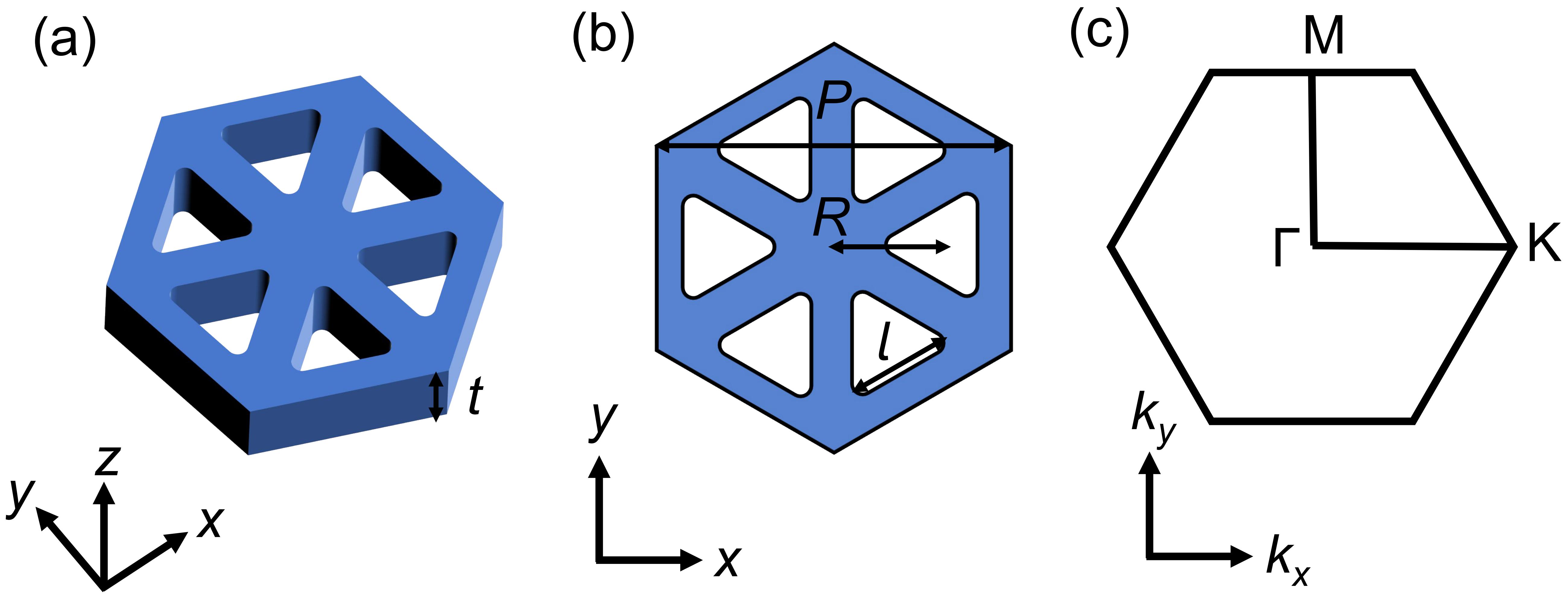}
\caption{(a, b) Schematic of a graphene-like SiN$_x$ PhCSs with a hexagonal lattice of etched triangular air holes. The PhCS is immersed in the air background. The thickness and lattice period of the PhCS are $t$=100 nm and $P$=496 nm, respectively. All triangular air holes have a side length of $l$=150 nm and a fillet of 25 nm. $R$ is the distance from the center of the triangular air hole to the center of the unit cell. (c) Brillouin zone of the PhCS in (a).}
\label{fig-1}
\end{figure}

The designed SiN$_x$ PhCS is shown in Fig.~\ref{fig-1}(a, b) with thickness of $t$=100 nm and lattice period of $P$=496 nm. $R$ is the distance from the center of the triangular air hole to the center of the unit cell. Fig.~\ref{fig-1}(c) is the Brillouin zone. Such configuration has been widely adopted to investigate the optical analogs of quantum spin Hall effects \cite{wu2015scheme,yang2018visualization,barik2018topological,parappurath2020direct,liu2020z2,li2021experimental,liu2020generation,guddala2021topological,arora2022breakdown}.
The topological properties of the bulk bands are determined by the ratio of $P/R$. Previous studies reveal that this system has trivial band topology for $P/R>3$ due to the frequency of the odd modes ($p$ orbitals) is lower than that of even modes ($d$ orbitals) at $\Gamma$ point, whereas the system has non-trivial band topology for $P/R<3$ due to the band inversion occurring between odd and even modes. It is an open system when the PhCSs are immersed in a nonopaque background medium. In such case, Bloch states around the $\Gamma$ point can have far-field radiation. Far-field properties of topological edge states aroused from band topology have been well studied \cite{barik2018topological,parappurath2020direct,liu2020z2,arora2022breakdown}. However, the radiation topology of this system has not been explored yet. Next, we mainly focus on the radiation topology of the system and explore the relationship between band topology and radiation topology.

\begin{figure*}
\centering{}\includegraphics[width=0.95\textwidth]{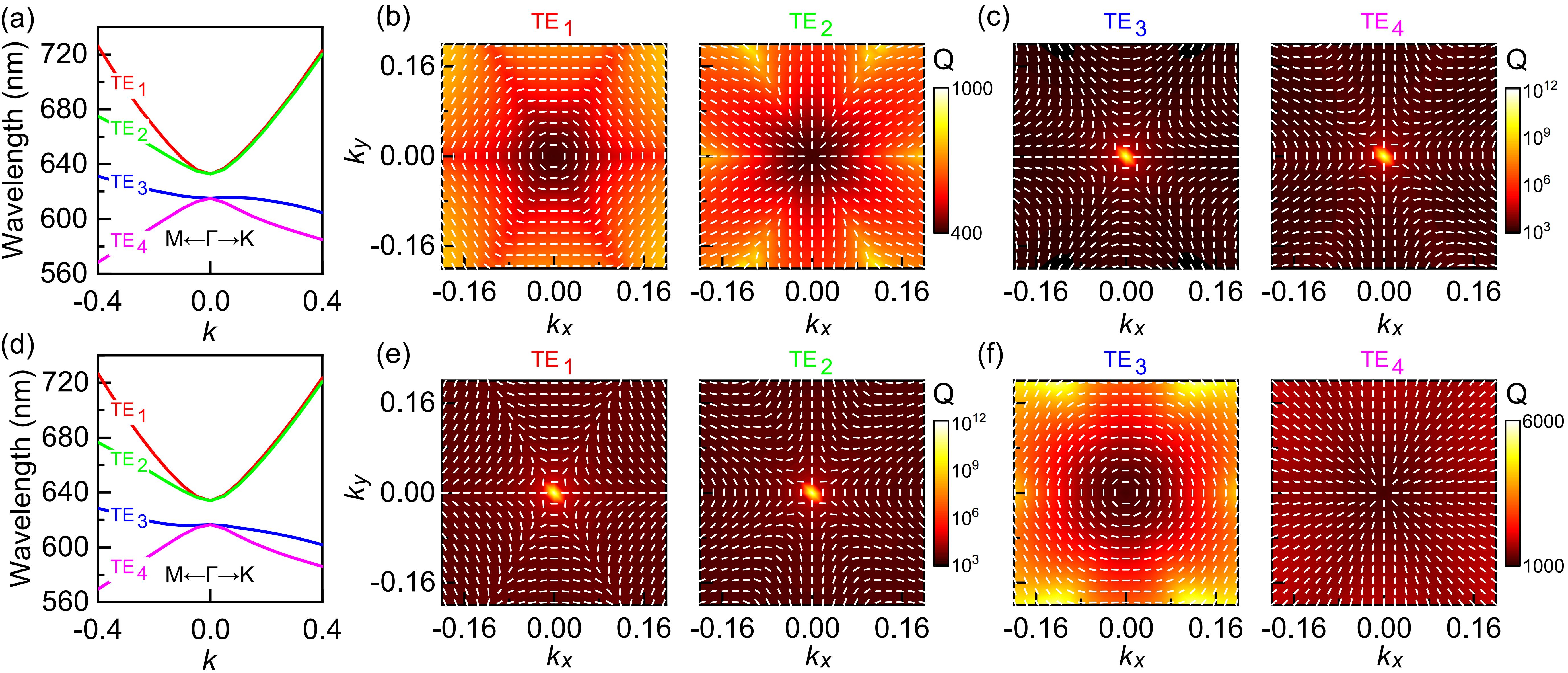}
\caption{(a, d) Calculated transverse electric (TE)-like band structure with $R=148$ nm in (a) and $R=175.5$ nm in (d). (b-c) and (e-f) are the calculated far-field polarization vectors (white lines) around the center of the Brillouin zone for the four photonic bands in (a) and (d), respectively. The distributions of $Q$ factor are served as background in (b-c) and (e-f). The units of $k$ and $k_x$ ($k_y$) are $2\pi /P$ and $\pi /P$ ($\pi /P$), respectively.}
\label{fig-2}
\end{figure*}

Fig.~\ref{fig-2}(a) is the band structure for $R=148$ nm. There are four transverse electric (TE)-like photonic bands in the considered wavelength range. It can be seen that the band topology is trivial from the previous analysis. It should be noticed that there is no band degeneracy above the light cone when off $\Gamma$ point and there are also no compatible diffraction channels of the radiating Bloch states. In such case, the far-field polarization vector [${\bm c}(\bm k_{\parallel})$] of the radiating Bloch states with Bloch wave vector $\bm k_{\parallel} \neq \bm 0$ can be obtained by
\begin{align}
  {\bm c}(\bm k_{\parallel})=\iint_{u.c.} {\bm E}(x, y, z)e^{i(k_x*x+k_y*y)}dxdy
\end{align}
The integration is calculated in one unit cell (u.c.) on an $xy$ plane outside the PhCSs. The state of polarization (SoP) can be described by Stokes parameters $(S_0, S_1, S_2, S_3)$ . Then, we project ${\bm c}(\bm k_{\parallel})$ into the $xy$ plane: ${\bm c}_{xy}(\bm k_{\parallel})= c_x(\bm k_{\parallel}) \hat{x}+c_y(\bm k_{\parallel}) \hat{y}$, where $c_x(\bm k_{\parallel})=\hat{x} \cdot \bm c(\bm k_{\parallel})$ and $ c_y(\bm k_{\parallel})=\hat{y} \cdot \bm c(\bm k_{\parallel})$. In such case, the Stokes parameters can be expressed as the following form,
\begin{equation}
\begin{aligned}
  S_0&=\vert c_x(\bm k_{\parallel})\vert ^2+\vert c_y(\bm k_{\parallel})\vert ^2\\
  S_1&=\vert c_x(\bm k_{\parallel})\vert ^2-\vert c_y(\bm k_{\parallel})\vert ^2\\
  S_2&=2Re[c_x^\ast(\bm k_{\parallel})c_y(\bm k_{\parallel})]\\
  S_3&=2Im[c_x^\ast(\bm k_{\parallel})c_y(\bm k_{\parallel})]
\end{aligned}
\end{equation}
The Stokes phase $\phi(\bm k_{\parallel})$ is expressed as,
\begin{equation} \label{St}
  \phi(\bm k_{\parallel})=\frac{1}{2}arg(S_1+iS_2)
\end{equation}
Then, one can define topological charge $q$ carried by the polarization singularity,
\begin{equation} \label{tq}
  q=\frac{1}{2\pi}\oint_L d\bm k_{\parallel}\cdot\nabla_{\bm k_{\parallel}}\phi(\bm k_{\parallel})
\end{equation}
where $L$ is a closed loop around the singular point of polarization in the counterclockwise direction. $q$ is equal to the winding number of ${\bm c}_{xy}(\bm k_{\parallel})$ around singular points for linear polarization.

The far-field polarizations of the radiating Bloch states for the band structure in Fig.~\ref{fig-2}(a) are shown in Fig.~\ref{fig-2}(b, c). Here, the ${\bm c}_{xy}(\bm k_{\parallel})$ is shown by line segments
without arrows due to one cannot distinguish ${\bm c}_{xy}(\bm k_{\parallel})$ and $-{\bm c}_{xy}(\bm k_{\parallel})$ under temporal harmonic oscillations. It can be seen from Fig.~\ref{fig-2}(b, c) that the far-field polarizations are linearly polarized due to the C$_{2}$ rotational symmetry and time reversal symmetry. The results in Fig.~\ref{fig-2}(b) show that the distributions of far-field polarizations for the TE$_1$ and TE$_2$ bands are similar to the azimuthally and radially linearly polarized vectorial optical fields (VOFs) in the real space, respectively. Thus, the distributions of ${\bm c}_{xy}(\bm k_{\parallel})$ create a polarization vortex in the momentum space and the center of the vortex is at $\Gamma$ point. It can be seen that the value of $q$ is equal to +1 for the TE$_1$ and TE$_2$ bands in Fig.~\ref{fig-2}(b) from the winding number of ${\bm c}_{xy}(\bm k_{\parallel})$ around $\Gamma$ point, which can also be confirmed by the corresponding Stokes phase $\phi(\bm k_{\parallel})$ in the supplementary materials of Fig.~S1.  Although the polarization distribution configurations in Fig.~\ref{fig-2}(b) are the same as BICs, the states at the V-point ($\Gamma$ point) are not BICs. This can be seen from the distributions of $Q$ factor in Fig.~\ref{fig-2}(b), which are both 412.6 at the V-point. Thus, the radiating V-point can be created when there exists band degeneracy in momentum space.

The distributions of the far-field polarizations in Fig.~\ref{fig-2}(c) are more complex than that in Fig.~\ref{fig-2}(b). $c_y(\bm k_{\parallel})$ is zero along the directions of $\Gamma K$ and $\Gamma M$ for the TE$_3$ band in Fig.~\ref{fig-2}(c). Based on distributions of ${\bm c}_{xy}(\bm k_{\parallel})$ and Stokes phase, the topological charge is $q=-2$ for the TE$_3$ band in Fig.~\ref{fig-2}(c). A similar analysis can be applied to the TE$_4$ band in Fig.~\ref{fig-2}(c), which shows that its topological charge is also -2. Unlike the results in Fig.~\ref{fig-2}(b), the $Q$ factor at $\Gamma$ point is infinite in Fig.~\ref{fig-2}(c). Thus, the states at the $\Gamma$ point are BICs for the TE$_3$ and TE$_4$ bands in Fig.~\ref{fig-2}(a). Although the band topology is trivial, the radiation topology is indeed non-trivial.

The photonic band structure for $R=175.5$ nm is shown in Fig.~\ref{fig-2}(d), whose dispersion curves are quite similar to that in Fig.~\ref{fig-2}(a). However, its band topology is non-trivial due to $P/R>3$. The distributions of the far-field polarizations are shown in Fig.~\ref{fig-2}(e, f) for the band structures in Fig.~\ref{fig-2}(d). Interestingly, the results in Fig.~\ref{fig-2}(e) show that the $q$ is -2 for the TE$_1$ and TE$_2$ bands, while the $q$ is +1 for the TE$_3$ and TE$_4$ bands as shown in Fig.~\ref{fig-2}(f). This means that when the band topology undergoes a phase transition, the radiation topology also undergoes a topological phase transition. Therefore, this result shows that although the connection between topological properties of band topology and radiation topology may not be obvious, there may be a correlation between their topological phase transition processes.

\begin{figure*}
\centering{}\includegraphics[width=0.8\textwidth]{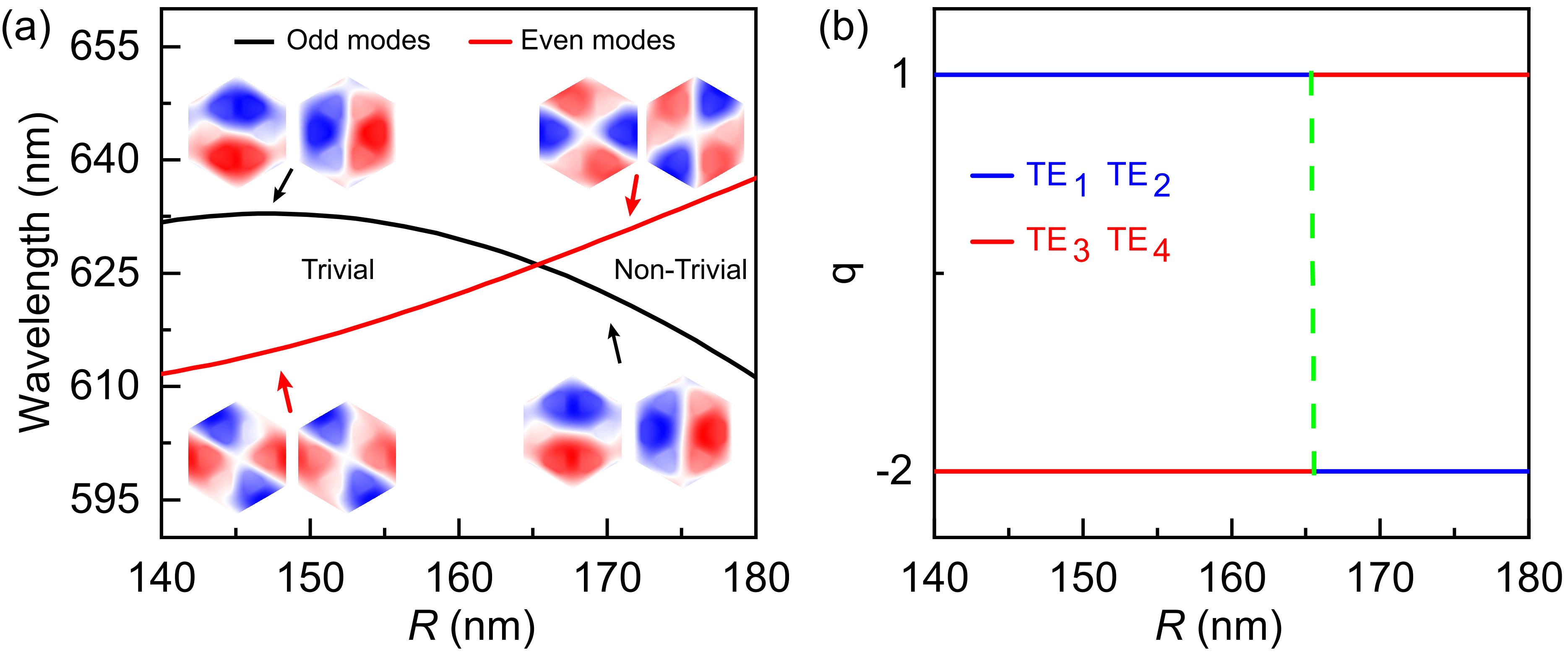}
\caption{(a) Illustration of band inversion process between odd and even modes at $\Gamma$ point with increasing $P$. The topological phase transition of band topology occurs when band inversion occurs. The insets in (a) are the field distributions of the odd modes and even modes at $\Gamma$ point for the $z$ component of the magnetic field. (b) The evolution of topological charge ($q$) with $R$.}
\label{fig-3}
\end{figure*}

To confirm this, we further study how their topological properties vary with $R$. As shown in Fig.~\ref{fig-3}(a), the topological phase transition of band topology occurs when $R$ changes from less than $P/3$ to greater than $P/3$ due to the band inversion. As shown in Fig.~\ref{fig-3}(b), when $R$ is less than $P/3$, the $q$ of the TE$_{1/2}$ and TE$_{3/4}$ Bands remains +1 and -2, respectively. However, once $R$ is larger than $P/3$, the $q$ of TE$_{1/2}$ (TE$_{3/4}$ ) bands changes from +1 to -2 (from -2 to +1). The phase transition critical points for both band topology and radiation topology are $R=P/3$, and their phase transitions are both caused by band inversion at the $\Gamma$ point. In fact, the $q$ is determined by the band representation (symmetry) of the Bloch mode at the high symmetry point \cite{zhen2014topological,zhang2018observation}. Thus, the $q$ does not change unless the band representation changes when the parameter changes do not change the symmetry of the system. Band representation changes often require band gap to close and reopen, leading to the phase transitions of band topology. Such observation suggests that one can indeed probe phase transitions of band topology using radiation topology.

It should be noted that before and after the phase transitions, the total topological charge of the radiation topology of the four bands remains unchanged. Thus, global topological charge conservation still holds during the phase transition process. Another important feature is that the band inversion mechanism can serve as an important route to explore the dynamics of topological polarization singularity and manipulate the SoP in the far-field, as shown in Fig.~\ref{fig-2} and Fig.~\ref{fig-3}. Specially, the change of $q$ after the band inversion is 3 for the four bands. The BICs also move from TE$_{3/4}$ to TE$_{1/2}$ bands after the band inversion. Moreover, the SoP near the polarization singularity is significantly affected by the topological charge. It can be seen from Fig.~\ref{fig-2} that when the band inversion occurs, the SoP of TE$_{1/2}$ and TE$_{3/4}$ bands in the momentum space are exchanged.

\begin{figure*}[h]
	\centering{}\includegraphics[width=0.93\textwidth]{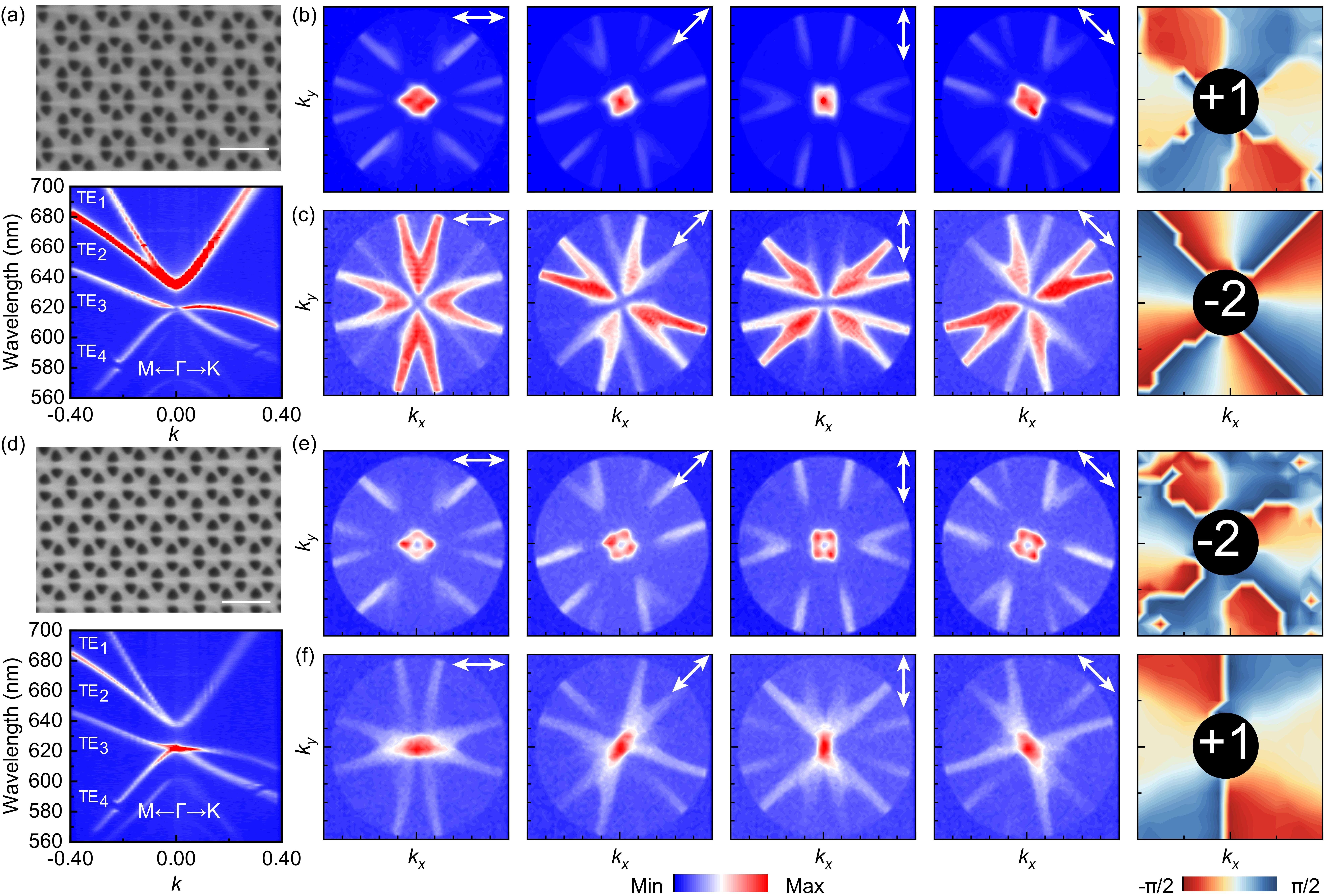}
	\caption{The SEM image and band dispersions of the band topologically trivial lattice R=148 nm in (a) and non-trivial lattice R=175.5 nm in (d). Polarization-resolved summed isofrequency contours and the corresponding Stokes phase maps of band topologically trivial lattice in (b, c) and non-trivial lattice in (e, f). The filtered center wavelengths in (b, e) and (c, f) are 640 nm and 620 nm, respectively. The 10 nm filter bandwidth in (b-c, e-f) is marked in the supplementary materials of Fig.~S2 (a, b). The scale bar of SEM images is 500 nm. White arrows in the isofrequency contours denote the direction of the linear polarizer. The units of $k$ and $k_x$ ($k_y$) are $2\pi /P$. The ranges of $k_x$ and $k_y$ are from -0.4 to 0.4 for isofrequency contours, and from -0.08 to 0.08 for Stokes phase maps. The number in the Stokes phase map is the topological charge of radiation topology.}
\label{fig-4}
\end{figure*}

We further fabricate the proposed band topologically trivial ($R=148$ nm) and non-trivial ($R=175.5$ nm) lattices based on SiN$_x$ PhCSs and test them by our homemade 4f spectroscopy system (See methods). The scanning electron microscopy (SEM) images and measured photonic dispersions for $R=148$ nm and $R=175.5$ nm samples are shown in Fig.~\ref{fig-4}(a) and Fig.~\ref{fig-4}(d), respectively. The plotted data are the intensity of photoluminescence of the SiN$_x$ PhCSs as a function of the in-plane wave vector and wavelength. It displays that the experimental results show good agreement with the calculation depicted in Fig.~\ref{fig-2}(a) and Fig.~\ref{fig-2}(d), except only with a slight wavelength difference.

Next, we experimentally demonstrate for the first time that one can probe the topological phase transitions of band topology via radiation topology. To manifest the radiation topology around $\Gamma$ point, we measure the polarization-resolved summed isofrequency contours with bandpass filters ($\sim$10 nm bandwidth) for two center wavelengths (CWL) of $\lambda$=620 and 640 nm. The measured isofrequency contours at four polarization angles of $0\degree$, $45\degree$, $90\degree$ and $135\degree$ are shown in Fig.~\ref{fig-4}(b, c) for $R=148$ nm and Fig.~\ref{fig-4}(e, f) for $R=175.5$ nm. The polarization angle is the orientation of the polarizer axis relative to the horizontal plane. With C$_2$ symmetry in our structure, the polarization states of these Bloch modes are almost linear \cite{zhen2014topological,kang2022merging}. When the far-field polarization azimuths of certain states are not perpendicular to the polarizer, the signal at such points will be transmitted, appearing as bright patterns. Thus, the pattern rotates along with the polarizer when the polarization vortex is present, which gives a vivid picture of the winding of far-field polarization vectors \cite{zhang2018observation,2019ChenObserving}. On the other hand, the quantitative method to determine the topological charge is to see the distributions of the Stokes phase $\phi(\bm k_{\parallel})$ defined by Eq.~\ref{St} in momentum space. Here, in experiments, the Stokes phase can be obtained by measuring the photoluminescence intensities at four polarization angles of $I_{0\degree}$, $I_{45\degree}$, $I_{90\degree}$ and $I_{135\degree}$ to determine the Stokes parameters $S_1=\frac{I_{0\degree}-I_{90\degree}}{I_{0\degree}+I_{90\degree}}$ and $S_2=\frac{I_{45\degree}-I_{135\degree}}{I_{45\degree}+I_{135\degree}}$ \cite{gbur2016singular}.

For the lattice with trivial band topology,, there is a bright spot around the $\Gamma$ point for the 640 nm-CWL filtered spectra in Fig.~\ref{fig-4}(b). It is because the band dispersions between TE$_{1}$ and TE$_{2}$ bands have tiny differences around the $\Gamma$ point and the far-field polarizations of the two bands are  orthogonal [see Fig.~\ref{fig-2}(a, b)]. However, we can see the far-field polarizations that spin with the rotation of the polarizer, as shown in Fig.~\ref{fig-4}(b), which indicates the existence of a polarization vortex for TE$_{1/2}$ band in Fig.~\ref{fig-4}(a). We also give the distributions of $\phi(\bm k_{\parallel})$ in momentum space. The Stokes phase in Fig.~\ref{fig-4}(b) experiences a $+2\pi$ change along a closed loop in the counter-clockwise direction, which clearly shows that TE$_{1/2}$ band in Fig.~\ref{fig-4}(a) has a polarization vortex at $\Gamma$ point, and $q$ is $+1$. The 620 nm-CWL filtered spectra have four clear lobes with a dark core in Fig.~\ref{fig-4}(c), because the dispersions of TE$_{3}$ and TE$_{4}$ bands are separated. In such a case, the Stokes phase accumulates $-4\pi$ counter-clockwise in a closed loop, compatibly with the situation of $q=-2$. The experimental results also confirm that the radiation topology of the band can be non-trivial when its band topology is trivial.

For the lattice with non-trivial band topology, similar analyses could apply to the results in Fig.~\ref{fig-4}(e) and Fig.~\ref{fig-4}(f). $q$ are -2 and +1 for Fig.~\ref{fig-4}(e) and Fig.~\ref{fig-4}(f), respectively. The topological charge around the $\Gamma$ point are swapped between TE$_{1/2}$ and TE$_{3/4}$ bands when the band topology undergoes a topological phase transition. These experimental results provide solid evidence that radiation topology is a more straightforward way to access the phase transitions of band topology.

Note that the experimental results also confirm that the band inversion mechanism can be used to control the SoP and quality factor. For example, the BICs at $\Gamma$ point move from TE$_{3/4}$ band to TE$_{1/2}$ band when the band inversion occurs, manifesting a swapping of brightness at the $\Gamma$ point [see Fig.~\ref{fig-4}(a, d)]. Comparing the results of Fig.~\ref{fig-4}(b) and Fig.~\ref{fig-4}(e) [Fig.~\ref{fig-4}(c) and Fig.~\ref{fig-4}(f)], the SoP is exchanged around $\Gamma$ point after occurring band inversion.

In summary, we theoretically analyze and experimentally verify that one can probe the phase transition of band topology via radiation topology. Compared to previous bulk-boundary correspondence principle that requires the formation of domain walls, the proposed radiation topology scheme measures the topological properties of band topology without domain walls. Moreover, such approach can also be applied to more general situations where the crystal has no full bandgap [see Fig.~\ref{fig-4}(a, d)] and topological states. Besides, since the polarization vortex is a topological configuration that is insensitive to the changes of the external environment, the radiation topology can provide a stable way to measure the phase transition of band topology. In addition, we also found that the band inversion of near-field Bloch states could provide a new way to manipulate far-field radiation. Our research not only provides an insightful understanding between band topology and radiation topology, but also will boost the development of topological photonics, bringing essential promotion to many key applications, including tunable topological charge vortex laser, polarization control, topological light emissions, etc.

\section{Methods}	
	\section*{Methods}	
	\subsubsection*{{Numerical Simulations}}
The numerical results are simulated by using eigenvalue solve of the finite element method (FEM,COMSOL Multiphysics). Periodic boundary conditions and perfect matching layers are applied in the in-plane dimensions $(x,y)$ and the out-of-plane dimension ($z$), respectively.
	
	\subsubsection*{{Sample Fabrication}}
The samples were fabricated based on commercial SiN$_x$ windows, of which the thickness is around 100 nm. Firstly, the SiN$_x$ windows were cleaned under an oxygen plasma atmosphere by reactive ion etching(RIE) to remove the residues. Secondly, the windows were spin-coated with a layer of positive-tone electron beam resist (poly(methyl methacrylate) (PMMA)), and $\sim$200 nm thick PMMA film was obtained at 4000 r.p.m for 60 s and then baked on a hotplate at 180$^{\circ}$C for 60 s. Thirdly, the PMMA was patterned by electron beam lithography, and then developed in a mixture of methyl isobutyl ketone(MIBK) and isopropyl alcohol(IPA) for 40 s (MIBK: IPA=1:3) and fixed in IPA for 30 s. Then, the periodically EBL-etched PMMA layer acts as a mask in the subsequent RIE process. Anisotropic etching of the SiN$_x$ was achieved using a mixture of CHF$_3$ and O$_2$. After ensuring the SiN$_x$ had been etched through, the PMMA mask was eventually removed by RIE using O$_2$.

  \subsubsection*{Optical Characterization}
The 4f system was based on the optical Fourier transformation method performed by the optical lenses to obtain the momentum-space information \cite{2012Wagner}. Contrary to the traditional angle-variable measurement, it could capture the spectra in a single acquisition with high resolution in both energy and momentum, allowing a fast-speed measurement. The sample was placed on the precision rotation mounts to set for the specific direction. For angle-resolved photoluminescence, a 405 nm continuous wave laser served as the excitation source. The excitation beam was focused by a 0.40 NA long working distance objective lens(Nikon) with average power $\sim$0.25 mW. The back focal plane of the objective was imaged onto the entrance slit of the spectrometer(iHR550, Horiba) equipped with a liquid-nitrogen-cooled CCD detector (Symphony II 1024$\times$256 BIDD). The vertically aligned entrance slit was set to 100 $\mu$m and fully opened for wavelength-resolved spectra and isofrequency contours, respectively.

\begin{suppinfo}
		This Supporting Information is available free of charge via the internet at http://pubs.acs.org.
	\end{suppinfo}

\section*{Competing interests}
  The authors declare no competing financial interest.

\section*{Acknowledgments}
	This work is supported by the National Key Research and Development Program of China under Grant No. 2021YFA1400700, the Strategic Priority Research Program of Chinese Academy of Sciences (CAS) under Grant Nos. XDB28000000, the Key Research Program of Frontier Sciences of CAS under Grant Nos. XDPB22, and the National Natural Science Foundation of China (Grant Nos. 11974386, 61975016 and 12204041), and the Science and Technology Project of Guangdong (Grant No. 2020B010190001) and Natural Science Foundation of Beijing Municipality (Grant Nos. Z190006 and 1212013).

\section*{Author contributions}
B. L. Liu, J. F. Li and Y. G. Yao supervised the project. C. Y. Ji, B. L. Liu and Y. G. Yao conceived the idea. C.Y. Ji carried out the numerical simulations. B. L. Liu, C. Z. Gu and G. Wang coordinated the experimental investigations. W. Z. Lan and P. Fu fabricated the samples. W. Z. Lan performed the optical measurements. W. Z. Lan, C. Y. Ji and B. L. Liu did the data analysis and wrote the manuscript. All authors contributed to the revision and discussion of the paper.

%\section*{Supporting Information}
%\subsection*{Experimental Far-field Polarization Vectors Near the $\Gamma$ Point}
%	\begin{figure*}[h]
%		\centering{}\includegraphics[width=0.95\textwidth]{Fig_S1}
%		\caption{Far-field polarization vectors map. The TE$_{1(2)}$(a) and TE$_{3(4)}$(b) bands have topological charges of $+1$ and $-2$ in the trivial lattice. In contrast, the TE$_{1(2)}$(a) and TE$_{3(4)}$(b) bands have topological charge of $-2$ and $+1$ in the non-trivial lattice. The topological charges have exchanged after the phase transitions of band topology.}
%		\label{fig-S1}
%	\end{figure*}
\bibliography{refs}

\end{document}